# Phyllotaxis-inspired Nanosieves with Multiplexed Orbital Angular Momentum

Zhongwei Jin[1,2], David Janoschka[3], Junhong Deng[4], Lin Ge[5], Pascal Dreher[3], Bettina Frank[6], Guangwei Hu[1], Jincheng Ni[1], Yuanjie Yang[7], Jing Li[8], Changyuan Yu[1,9], Dangyuan Lei[10], Guixin Li[4], Shumin Xiao[11], Shengtao Mei[1], Harald Giessen[6]*, Frank Meyer zu Heringdorf[3]*, Cheng-Wei Qiu[1]*

[1]Department of Electrical and Computer Engineering, National University of Singapore,

4 Engineering Drive 3, Singapore 117583, Singapore

[2]College of Optical and Electronic Technology, China Jiliang University, Hangzhou 310018, China

[3]Faculty of Physics and Center for Nanointegration Duisburg–Essen (CENIDE), University of Duisburg–Essen,

Lotharstrasse 1-21, 47057 Duisburg, Germany

[4]Department of Materials Science and Engineering, Shenzhen Institute for Quantum Science and Engineering, Southern University of Science and Technology, Shenzhen 518055

[5]Beijing Qianjunyide technology co. Ltd., Beijing 100031, China

[6]Physics Institute and Stuttgart Center of Photonics Engineering (SCoPE), University of Stuttgart, D-70569 Stuttgart, Germany.

[7]School of Physics, University of Electronic, Science and Technology of China, Chengdu 611731, China

[8]Key Laboratory of Optoelectronic Materials, Technical Institute of Physics and Chemistry, Chinese Academy of Sciences, Beijing 100190, China

[9]Department of Electronic and Information Engineering, The Hong Kong Polytechnic University, Hung Hom,

Kowloon, Hong Kong SAR, China.

[10]Department of Materials Science and Engineering, City University of Hong Kong, 83 Tat Chee Avenue,

Kowloon, Hong Kong SAR, China

[11]State Key Laboratory on Tunable Laser Technology, Ministry of Industry and Information Technology Key Lab

of Micro-Nano Optoelectronic Information System, Shenzhen Graduate School, Harbin Institute of Technology,

Shenzhen 518055, China

[12]SZU-NUS Collaborative Innovation Center for Optoelectronic Science and Technology, Shenzhen University,

Shenzhen 518060, China

*Corresponding authors: *giessen@pi4.uni-stuttgart.de; meyerzh@uni-due.de; chengwei.qiu@nus.edu.sg;*

## Abstract

Nanophotonic platforms such as metasurfaces, achieving arbitrary phase profiles within ultrathin thickness, emerge as miniaturized, ultracompact and kaleidoscopic optical vortex generators. However, it is often required to segment or interleave independent sub-array metasurfaces to multiplex optical vortices in a single nano-device, which in turn affects the device's compactness and channel capacity. Here, inspired by phyllotaxis patterns in pine cones and sunflowers, we theoretically prove and experimentally report that multiple optical vortices can be produced in a single compact phyllotaxis nanosieve, both in free space and on a chip, where one meta-atom may contribute to many vortices simultaneously. The time-resolved dynamics of on-chip interference wavefronts between multiple plasmonic vortices was revealed by ultrafast time-resolved photoemission electron microscopy. Our nature-inspired optical vortex generator would facilitate various vortex-related optical applications, including structured wavefront shaping, free-space and plasmonic vortices, and high-capacity information metaphotonics.

## Keywords

phyllotaxis vortex nanosieve, free-space, plasmonic vortices, ultrafast time-resolved

## Introduction

The vortex is ubiquitous in nature, spanning from the galaxy, ocean flow, to phyllotaxis. In electromagnetic wave, the vortex can be found in spiral wavefront profiles, representing the orbital angular momentum (OAM) of light[1-5]. Analogous to the natural vortex, such as tornado and whirlpool, the optical vortices (OVs) carrying OAM have helical phase fronts and donut-shaped intensity profiles[6-10]. The last few decades have witnessed the prosperous development in the advanced application of OVs, such as in quantum optical communications[11], particle manipulation[12-14], super-resolution imaging[15], and multi-channel information storage[16-18]. So far, various kinds of optical components have been proposed to generate OAM beams, such as spiral phase plates[19], forked holograms[20], metasurfaces[21-23], and chip-scale microlasers[24-28]. Recently, multifunctional metasurfaces[18,21,22,29,30] have been investigated to generate multiple OVs within one nano-device. Such metasurfaces often resort to segmented or interleaved sub-array meta-atoms to multiplex OAMs in a single device. As a result, each inclusion seems only responsible for one specific topological charge. Meanwhile, sufficient distance is required between meta-atoms to suppress cross coupling, which in turn degrades the device's compactness and channel capacity. To deal with this problem, one may search therapies/solutions from either frequency domain or space domain. Previously, Elhanan Maguid et. al[21] introduced asymmetric harmonic response geometric phase metasurface which realized OAM multiplexing through superposition of different harmonic components in the momentum space. In this work, we realize both free-space and near-field OAM multiplexing based on structure degeneracy in the space domain.

## Results

Intriguingly, planar spirals such as Archimedean spirals[31], logarithmic spirals[32] and Fermat spirals[33] can generate photonic OAMs with helical phase fronts. Among the Fermat spirals, the Vogel spiral[34], also known as the "golden ratio" spiral, has been frequently studied for its unique growing pattern[35-37]. The pattern of a Vogel spiral[34] in polar coordinates can be described as $r = c\sqrt{n}$, $\phi = n \cdot 137.5°$. Here, $n$ is the ordering number of a floret, $c$ is a scaling constant, $r$ is the radial distance between the $n^{th}$ floret and the center of the capitulum, $\phi$ is the angle between the reference direction and the position vector of the $n^{th}$ floret, and 137.5° is the "golden angle". The Vogel spiral is well known as one of the phyllotaxis geometries in nature, which exists in many plants including pine cones, sunflower seeds, and so on[35]. As shown in Figure 1a, multiple sets of clockwise and anti-clockwise spirals can be encoded from such a phyllotaxis geometry pattern. And the numbers of spiral arms contained in different sets are in coincidence with the Fibonacci numbers.

Such interesting phenomenon naturally arouses our interest to investigate the link between the nature-inspired pattern and optical vortices. In order to solve this puzzle, we simulated the diffraction pattern of a "golden-ratio" Vogel spiral nanosieve ($c = 2.5$, $n$ starts from 1, ends at 936) in its Fresnel region at $z = 300\mu m$ upon 633nm light's incidence. Indeed, the OAM spectrum analysis[38] reveals that the diffracted pattern of such a mask contains a series of OAM modes (Figure 1a), in coincidence with the numbers of spiral arms which can be encoded from the pattern. Hence, we infer that phyllotaxis-alike patterns concealing multiple spiral structures may enable the creation and multiplexing of OVs. Such beauty in nature inspires us to design phyllotaxis-alike nanosieves which can generate beams containing multiple OAM modes for both free-space and on-chip optical systems.

**Design concept of phyllotaxis-inspired nanosieves**

First, we revisit the "vortex comb" phenomenon[39] to obtain the working principle of our phyllotaxis-inspired nanosieves and extend it to both free-space and on-chip optical systems. Under such circumstances, light emitted from each subwavelength nanohole of nanosieves can be approximated as a point source[40,41]. Considering a total of $M$ point sources arranged along the azimuthal domain with equal angular separation, light emitted from the $a^{th}$ nanohole at plane-wave incidence can be decomposed into the summation of a set of orthogonal LG modes[42]:

$$\phi\left(\rho, \frac{2\pi \cdot a}{M}, z\right) \propto \sum_{p=0}^{\infty} \sum_{l=-\infty}^{\infty} c_{p,l} I_{p,l}(\rho, z) exp\left[-il\left(\frac{2\pi \cdot a}{M}\right)\right] \cdot exp\left(i\sigma \frac{2\pi \cdot a}{M}\right) \quad (1)$$

In equation (1), the LG modes are written in the form of $I_{p,l}(\rho, z)exp(il\theta)$, where $I_{p,l}(\rho)$ denotes the complex amplitude of the corresponding LG mode, and $c_{p,l}$ denotes the expanded coefficient. $\rho$ denotes the radial distance of the point source to the original point and $z$ represents the focal distance. In free space, $z > 0$, and $\sigma = 0$ as long as the incidence is plane wave; while in the on-chip optical system, $z$ can be approximated as zero, and $\sigma = \pm 1$ for right- (RCP) and left-handed circularly polarization (LCP) states, owing to the appropriate spin-to-orbital conversion mechanism. Therefore, the final field distribution can be approximated as the interference of such $M$ point sources, which is given by the summation of these individual elementary waves:

$$I = \sum_{a=0}^{M-1} exp\left[-i(l-\sigma)\frac{2\pi \cdot a}{M}\right] \cdot \sum_{p=0}^{\infty} \sum_{l=-\infty}^{\infty} c_{p,l} I_{p,l}(\rho, z) \quad (2)$$

Since $\sum_{a=0}^{M-1} exp\left[-i(l-\sigma)\frac{2\pi \cdot a}{M}\right]$ is the summation of a finite geometric series and can be easily calculated as:

$$\sum_{a=0}^{M-1} exp\left[-i(l-\sigma)\frac{2\pi \cdot a}{M}\right] = \frac{1-exp\left[-i(l-\sigma)\frac{2\pi \cdot aM}{M}\right]}{1-exp\left[-i(l-\sigma)\frac{2\pi \cdot a}{M}\right]} = \begin{cases} M, & l = NM + \sigma \\ 0, & l \neq NM + \sigma \end{cases} \quad (3)$$

Combining equations (2) and (3), we can conclude the following: For the free-space optical system, the interference pattern of such $M$ point sources can be expressed as:

$$I = \begin{cases} M \cdot \sum_{p=0}^{\infty} \sum_{l=-\infty}^{\infty} c_{p,l} I_{p,l}(\rho, z), l = NM \\ 0, l \neq NM \end{cases}, \quad (4)$$

while for an on-chip optical system, the interference pattern of such $M$ point sources can be expressed as:

$$I = \begin{cases} M \cdot \sum_{p=0}^{\infty} \sum_{l=-\infty}^{\infty} c_{p,l} I_{p,l}(\rho), l = NM + \sigma \\ 0, l \neq NM \end{cases}, \quad (5)$$

where $N = 0, \pm 1, \pm 2, \pm 3, ...$ in equations (3), (4), and (5).

In brief, we remark two important conclusions from the above theoretical discussion. First, multiple orders of OAM modes can be generated both in free space and in the near field via a single nanosieve device, as visible in equations (4) and (5). The appearance of these sequential OAM orders is deeply rooted in the rearrangement of the nanoholes into different sets of spirals (see various colored spiral lines in Fig. 1a and more details in the following designs), and those spirals will render the corresponding different OAMs. This is also the reason of the emerged Fibonacci sequential OAM orders embedded in a "golden-ratio" phyllotaxis nanosieve, inspired of which we call our compact devices *phyllotaxis-inspired vortex nanosieves*. Second, in free space, the OAM orders are independent of incident spins; while in the on-chip optical system, we can get a series of OAM modes containing spin-to-orbit conversion. Intrinsically, the surface plasmon polariton (SPP) wave excited via the circular-shape nanohole by circularly polarized light has different initial phases along different propagating directions. However, in the on-chip optical system, only SPP wave propagating towards the center of the nanosieve will interfere and form the vortices. Therefore, under circularly polarization illumination, our *phyllotaxis-inspired vortex nanosieve* realizes spin-to-orbit conversion.

**Free-space phyllotaxis-inspired vortex nanosieve**

We employed the Fermat spiral with the formulation[33] $r_\theta = \sqrt{r_0^2 + 2lz_0\lambda \cdot \frac{\theta}{2\pi}}$ , ($r_0 \ll z_0$), to generate a beam with tailored OAM modes in the free-space optical system. Here, $\theta$ denotes the azimuthal angle of the spiral, $r_\theta$ denotes the spiral radius corresponding to azimuthal angle $\theta$, and $r_0$ is the starting radius of the spiral. Therefore, light penetrating the spiral slit structure will form a helical wavefront and accumulate a $l \cdot 2\pi$ phase difference on the designed focal distance $z_0$. Combining our previous derivation along with inspiration from the "golden ratio" phyllotaxis nanosieve, we repeated such spiral structure equally along the azimuthal angular domain *l* times and segmented the spiral slit structure into azimuthal equally separated nanoholes to obtain a *phyllotaxis-inspired vortex nanosieve*. Specifically, we choose $\lambda = 633\text{nm}, r_0 = 22\mu\text{m}, z_0 = 250\mu\text{m}$ and $l = 13$. Here, we vary spiral azimuthal angle covering from 0 to $3\pi$. Each of the 13 spirals is azimuthal equally segmented into 72 nanoholes. Hence, our *phyllotaxis-inspired vortex nanosieve* is composed of 936 arranged nanoholes in total (Figure 1b).

As indicated by our theoretical insight, we now prove that our *phyllotaxis-inspired vortex nanosieve* can generate multiple OAMs beyond the topological charge of *l*. As the location of each nanohole is fixed, we can re-unite or re-sample the nanohole arrays. If we "string" the neighboring nanoholes following different trajectories, different spiral patterns can be encoded. As is shown in the right panel of Figure 1b, four obvious sets of motifs can be encoded from the free-space *phyllotaxis-inspired vortex nanosieve*, which are 13 clockwise spirals, 39 anti-clockwise spirals, 52 clockwise spirals, and 91 anti-clockwise spirals correspondingly. Based on this, we can infer that light coming from our free-space *phyllotaxis-inspired vortex nanosieve* will simultaneously carry four OAM modes with different helical wavefronts, with the correspond topological charge of *l*=+13, -39, +52, and -91, whose numerical amplitude intensity profile is also shown the right panel of Figure 1b.

To verify our analysis, both numerical simulation and experiments were carried out. Figure 2a and 2b shows the simulated intensity and phase profiles of the diffraction pattern of the free-space *phyllotaxis-inspired vortex nanosieve* at $z=250$ μm upon 633 nm's illuminance respectively. It can be clearly observed that the generated on-axis four OAM patterns in Figure 2a are the superposition of the four simulated modes shown in Figure 1b with nearly no distortion. Meanwhile, one can directly obtain the modes' information from the corresponding phase profile. The free-space *phyllotaxis-inspired vortex nanosieve* sample was fabricated using focused-ion beam (FIB) technique on a 120-nm thick Au film above a glass substrate. The radius of each milled nanohole is 1μm. Figure 2c shows the top-view scanning electronic microscopic (SEM) picture of the sample. The measured intensity profiles by light beams with different wavelengths (experimental setup can be found in Supporting Information Fig.S.1) are shown in Figure 2d. As mentioned, the designed focal plane is 250 μm above the nanosieve for 633 nm incident light. According to Fresnel's principle, the focal planes would change to 298 μm and 356 μm for the incident wavelength of 532 nm and 445 nm, respectively. All the intensity profiles are captured near the corresponding focal planes of the nanosieve. It is interesting to know that the OAM mode sequence obtained from our free-space *phyllotaxis-inspired vortex nanosieve* is actually the first four numbers in the Lucas numbers sequence multiply by 13, which are 1, 3, 4 and 7. The Lucas numbers are an integer sequence that are closely related to the more well-known Fibonacci numbers, and are obtained like the Fibonacci series, but with starting values 2 and 1(2, 1, 3, 4, 7, 11,…). Statistics showed that 4% of the patterns of pine trees grown in Norway follow the Lucas numbers, while the majority of the rest follow Fibonacci numbers[43].

**Plasmonic phyllotaxis-inspired vortex nanosieve**

For the design of the plasmonic *phyllotaxis-inspired vortex nanosieve*, we employed a spiral structure with the formulation[44] $r_l(\theta) = r_0 + \frac{\theta \cdot l}{2\pi} \cdot \lambda_{\text{SPP}}$. Here, $l$ is the designed topological charge; $\theta$ is the azimuthal angle; $r_0$ denotes the initial radius of the spiral; $r_l(\theta)$ denotes the spiral radius corresponding to azimuthal angle $\theta$ associated with a topological charge of $l$. Besides, $\lambda_{SPP}$ denotes the SPP wavelength which is around 606 nm at the interface of gold and air at the pump laser frequency of 633nm. In our design, we specifically set $l = 40$, $r_0 =$

10μm, and $\theta$ from 0 to $\frac{5\pi}{l}$. To construct the *phyllotaxis-inspired vortex nanosieve*, 40 such spirals are azimuthal equally arranged, each segmented into 4 nanoholes (Figure 3a).

As shown in Figure 3a, such an on-chip *phyllotaxis-inspired vortex nanosieve* could encode 40 anti-clockwise rotated spirals (red solid circles and dotted lines in Figure 3a) and 80 anti-clockwise rotated spirals (blue solid circles and dotted lines in Figure 3a). Thus, upon excitation, SPPs excited at each nanohole propagate towards the center, and at least two distinct plasmonic vortex modes would emerge. Considering the spin-orbit conversion[45] suggested by equation (5), for right-handed circular polarization (RCP), the generated plasmonic vortex modes should be -79, -39, and +1; while for the left-handed circular polarization (LCP), the resultant plasmonic vortex modes would be -81, -41, and -1. Note that the vortex modes with $l = \pm 1$ (positive for RCP and negative for LCP incidence) stem from the rearrangement of nanosieve into a circle with the equal radial distance away from the center. Such vortex can also be viewed as a deuterogenic plasmonic vortex[46].

In order to obtain accurate optical responses of the plasmonic *phyllotaxis-inspired vortex nanosieve*, both full-wave numerical simulations based on Lumerical FDTD and near-field measurement using scanning near-field optical microscope (SNOM, Ntegra solaris from NT-MDT Spectrum Instrument, Moscow, Russia) have been carried out. Detailed configurations for both simulation and measurement are provided in Supporting Information Figure S.2. The sample was fabricated using FIB technique through etching 120nm-thick Au nanoholes on top of glass substrate and the top-view SEM picture of the sample is provided in Figure 3b. The radius of each nanohole is 150nm. Figures 3c and 3d depict numerical intensity and phase profiles under opposite circular polarizations illumination and linear polarization illumination. According to the simulated intensity profiles in Figure 3c, three on-axis plasmonic vortex modes can be clearly observed, which agrees with our theoretical expectations. Meanwhile, the phase variations shown in Figure 3d further verify the results. Under RCP illumination, a $2\pi$ phase change in the center of the phase profile can be clearly observed, which denotes a plasmonic vortex with $l = +1$. Besides, $-39 \cdot 2\pi$ phase variation and $-79 \cdot 2\pi$ phase variation can be noticed in the corresponding regions of the phase profile, indicating plasmonic vortices with $l$ = -39 and $l$ = -79 respectively. Under LCP illumination, $-2\pi$, $-41 \cdot 2\pi$ and

$-81 \cdot 2\pi$ phase changes can be observed in the corresponding regions of the phase profile, indicating plasmonic vortices with $l$ = -1, -41 and -81. For linearly polarization (LP) incidence, the generated modes are a combination of the modes generated under RCP and LCP incidences as expected. This is because LP can be regarded as linear combination of RCP and LCP and our system is linear. The mode decomposition analysis in Supporting Information Figure S.3 also verifies this. These results are further supported by the experimental results as the measured intensity profiles (Figure 3e) agree with Figure 3c, which further justifies the effectiveness of our proposed *phyllotaxis-inspired vortex nanosieve*.

**Time-resolved investigation of the plasmonic phyllotaxis-inspired vortex nanosieve**

Note that the results above are the time-averaged field intensity distributions. We thus resort to time-resolved two-photon photoemission electron microscopy[47-50] (TR-PEEM) to analyze the spatiotemporal dynamic processes to reveal the fundamentals of our plasmonic phyllotaxis-inspired vortex nanosieve. The TR-PEEM process begins with a pump-probe excitation of surface plasmons from nanoholes etched in a single-crystal gold flake, then the interference of the propagating surface plasmon with a probe pulse (Figure 4a), finally the imaging of the ejected photoelectrons in a photoemission electron microscope (PEEM)[50].

Since the TR-PEEM is combined with a Ti:Sapphire laser system operating at 800 nm central wavelength, the plasmonic wavelength changes to $\lambda spp \approx 780$ nm and we adjusted our design accordingly. An top-view SEM image of the sample is provided in Figure 4b. We tested the sample using the TR-PEEM system under both RCP and LCP incidences, and the relevant videos are uploaded as Supplementary Information. We have summarized the snapshots from the TR-PEEM results under RCP incidence which represent the three main stages of the vortex in dynamics formation, which are formation, revolution, and decay. Figure 4c~4e are the raw data from the TR-PEEM results. Figure 4f~4h show processed images that correspond to the delay times used in Fig. 4c~4e. Using temporal Fourier filtering at the 1ω component removes the time independent static backgrounds and leaves us exclusively with the SPP dynamics[51]. The spatiotemporal investigation shows how the excited SPPs propagate both inward and outward along the radial coordinate of the nanosieve. While the outward propagating SPPs finally leave the field of view, the inward propagating SPPs interfere and form the vortices. According to the handedness of the spiraling wavefronts, we can identify three major stages of vortices' dynamics. In Figure 4c, f, the converging spiraling wavefronts begin to form the two plasmonic vortices. The handedness of the spiraling wavefronts is indicated by the black dashed arrows in Figure 4c, f, and agrees with the handedness of the two sets of spirals that comprise our plasmonic *phyllotaxis-inspired vortex nanosieve*. Subsequently, the resulting revolution of the vortices is depicted in Figure 4d, g, where inward and outward counter-propagating SP waves interfere and form radially standing but azimuthally rotating vortex fields. Finally, the vortices decay and dissolve, forming outward-propagating spiraling wavefronts as displayed in Figure 4e, h. Compared to the formation stage, the spiraling fringes of the two decaying vortices show inverted handedness.

In the TR-PEEM experiment, the plasmonic vortices are imaged using a pump-probe technique, where both the pump and the probe pulse are circularly polarized. The measured plasmonic vortices in the TR-PEEM experiment thus do not contain the helicity of the pump light[48]. Accordingly, in Figure 4f~4h, the number of lobes of the corresponding plasmonic vortices are 80 and 40 respectively, cancelling the effect of the σ term in Equation 5. As a result, the plasmonic vortex induced by the pure "spin-orbit conversion" phenomenon is also offset in the measured result. Instead of forming a small "circle" in the center of the measured profiles, a small solid "dot" emerges as the interference result.

**Conclusions**

In conclusion, inspired by the "golden-ratio" spiral phyllotaxis-inspired pattern in nature, we presented the idea of using *phyllotaxis-inspired vortex nanosieves* to generate optical beams carrying multi-mode vortices both in free space and in a plasmonic near field. Besides theoretical and numerical verifications, we designed and fabricated several *phyllotaxis-inspired vortex nanosieves* to realize multi-mode vortex manipulation. The phenomenon of multiplexed OAMs in our *phyllotaxis-inspired vortex nanosieves* comes from the embedded multiple spirals in a single device, each spiral set carrying a different vortex mode, which is confirmed by both steady-state and dynamic-state measured results. Such strategy offers a different recipe with multimode OAM manipulation for promising application such as on-chip photonic devices[52], optical communication, and even quantum chiral optics.

## Methods

**Numerical simulation**.

The diffractive patterns (intensity and phase profiles) for free-space *phyllotaxis-inspired vortex nanosieve* and the mode decomposition analysis for both free-space and on-chip *phyllotaxis-inspired vortex nanosieve* were performed by MATLAB. The simulated static characterization of on-chip *phyllotaxis-inspired vortex nanosieve* (intensity and phase profiles) was performed by Lumerical FDTD Solution (a commercial software), and the detailed simulation setup is provided in Supporting Material Part.2.

**Sample fabrication**.

Both the samples used for free-space and static on-chip characterization were fabricated on two pieces of 120-nm-thick Au film, respectively. The Au film was deposited on cleaned glass substrate using E-beam evaporation (HHV, AUTO500) with a deposited rate of 1 Å/s. Then the pattern for free-space and static on-chip characterization was sculptured on Au film using focused ion beam technique (FEI, Helios NanoLab 600i), which is controlled by the software named NanoBuilder. The current and energy of ion beam are 80 pA and 30 kV, respectively. And the dwell time and volume per dose of patterning conditions are 1 $\mu s$ and 0.15 $\mu m^2/nC$ in Nanobuilder, respectively.

For TR-PEEM measurements, we chemically synthesize single crystalline gold platelets[50] on n-doped silicon substrates, which have lateral dimensions of up to one hundred micrometers. We then structure these atomically flat surfaces with phyllotaxis-inspired patterns, using a focused beam of Au++ ions, which is generated by a Raith IonLine plus system.

**Sample characterization.**

The SNOM system used for static on-chip characterization was Ntegra solaris, NT-MDT Spectrum Instrument, Moscow, Russia. The Fiber SNOM probe was NF113NTF, Scansens GmbH, Ostec Instrument, Germany.

The TR-PEEM system used for investigating the temporal dynamics of the nanosieve was built around a SPE-LEEM III (ELMITEC Elektornenmikroskopie GmbH,Germany) that was combined with a Ti:Sapphire femtosecond laser oscillator (Femtolasers). Pump- and probe pulses were created and mutualy delayed in a home-built Pancharatnam's phase stabilized Mach-Zehnder Interferometer. We work in normal-incidence geometry, as described in Refs. [31,53]

## List of abbreviations

| | |
|---|---|
| **OAM** | orbital angular momentum |
| **OVs** | optical vortices |
| **SPP** | surface plasmon polariton |
| **SEM** | scanning electronic microscopy |
| **RCP** | right-handed circular polarization |
| **LCP** | left-handed circular polarization |
| **LP** | linearly polarization |
| **SNOM** | scanning near-field optical microscope |
| **TR-PEEM** | time-resolved two-photon photoemission electron microscopy |
| **FDTD** | finite-difference time-domain |

## Declarations

**Availability of data and materials**

The datasets used and/or analysed during the current study are available from the corresponding author on reasonable request.

**Competing interests**

The authors declare that they have no competing interests.

**Funding.**

This work was supported by the National Research Foundation, Prime Minister's Office, Singapore under Competitive Research Program Award NRF-CRP22-2019-0006; the Deutsche Forschungsgemeinschaft (DFG, German Research Foundation) –Project-ID 278162697–SFB 1242; and ERC Advanced Grant Complex Plan, BMBF, DFG and BW-Stiftung; and the Research Grants Council of Hong Kong (CRF Grant No. C6013-18G) and the City University of Hong Kong (Project No. 9610434); and the support from A*STAR under its AME YIRG Grant (Award No. A2084c0172).

**Authors' contributions**

Z.J. and C.W.Q. conceived the idea. Z.J., C.W.Q., S.M. and S.X. designed the nano-devices and did the optical characterization of the free-space phyllotaxis-inspired vortex nanosieve. J.D. and G.L. did the nanofabrication and took SEM pictures for both free-space phyllotaxis-inspired vortex nanosieve sample and the sample used in the static characterization of the on-chip phyllotaxis-inspired vortex nanosieve. G.L., J.L., D.L. and X.H did the static optical characterization of the on-chip phyllotaxis-inspired vortex nanosieve. D.J and F.M.H. did the dynamic optical characterization of the on-chip phyllotaxis-inspired vortex nanosieve. B.F. and H.G. did the nanofabrication of the sample used in dynamic characterization of the phyllotaxis-inspired vortex nanosieve. P.D. did the data analyzation of the dynamic characterization of the on-chip phyllotaxis-inspired vortex nanosieve. G.H., J.N., and C.Y. participated in the discussions and contributed valuable suggestions for the phyllotaxis-inspired vortex nanosieves. The paper was written by Z.J. with inputs from D.J., J.D., L.G., P.D., B.F., H.G., G.H., J.N., and C.W.Q. C.W.Q supervised the project. All authors analyzed the data, read and corrected the paper before paper submission. All authors read and approved the final manuscript.

**Acknowledgements**

Not applicable.

# References

1      Sun, J., Zeng, J., Wang, X., Cartwright, A. N. & Litchinitser, N. M. Concealing with Structured Light. *Scientific Reports* **4**, 4093, doi:10.1038/srep04093 (2014).

2      Du, L., Yang, A., Zayats, A. V. & Yuan, X. Deep-subwavelength features of photonic skyrmions in a confined electromagnetic field with orbital angular momentum. *Nature Physics* **15**, 650-654, doi:10.1038/s41567-019-0487-7 (2019).

3      Shen, Y. *et al.* Optical vortices 30 years on: OAM manipulation from topological charge to multiple singularities. *Light: Science & Applications* **8**, 90, doi:10.1038/s41377-019-0194-2 (2019).

4      Pan, J. *et al.* Index-Tunable Structured-Light Beams from a Laser with an Intracavity Astigmatic Mode Converter. *Physical Review Applied* **14**, 044048, doi:10.1103/PhysRevApplied.14.044048 (2020).

5      Lin, F., Qiu, X., Zhang, W. & Chen, L. Seeing infrared optical vortex arrays with a nonlinear spiral phase filter. *Opt. Lett.* **44**, 2298-2301, doi:10.1364/OL.44.002298 (2019).

6      Bliokh, K. Y., Rodriguez-Fortuno, F. J., Nori, F. & Zayats, A. V. Spin-orbit interactions of light. *Nat. Photon.* **9**, 796-808, doi:10.1038/nphoton.2015.201 (2015).

7      Wang, B. *et al.* Generating optical vortex beams by momentum-space polarization vortices

centred at bound states in the continuum. *Nature Photonics* **14**, 623-628, doi:10.1038/s41566-020-0658-1 (2020).

8   Tang, Y. *et al.* Harmonic spin–orbit angular momentum cascade in nonlinear optical crystals. *Nature Photonics* **14**, 658-662, doi:10.1038/s41566-020-0691-0 (2020).

9   Dorney, K. M. *et al.* Controlling the polarization and vortex charge of attosecond high-harmonic beams via simultaneous spin–orbit momentum conservation. *Nature Photonics* **13**, 123-130, doi:10.1038/s41566-018-0304-3 (2019).

10  Ji, Z. *et al.* Photocurrent detection of the orbital angular momentum of light. *Science* **368**, 763, doi:10.1126/science.aba9192 (2020).

11  Nagali, E. *et al.* Quantum Information Transfer from Spin to Orbital Angular Momentum of Photons. *Physical Review Letters* **103**, 013601, doi:10.1103/PhysRevLett.103.013601 (2009).

12  Mei, S. *et al.* Evanescent vortex: Optical subwavelength spanner. *Applied Physics Letters* **109**, 191107, doi:10.1063/1.4967745 (2016).

13  Padgett, M. & Bowman, R. Tweezers with a twist. *Nature Photonics* **5**, 343-348, doi:10.1038/nphoton.2011.81 (2011).

14  Shvedov, V. G. *et al.* Optical vortex beams for trapping and transport of particles in air. *Applied Physics A* **100**, 327-331, doi:10.1007/s00339-010-5860-4 (2010).

15  Kozawa, Y., Matsunaga, D. & Sato, S. Superresolution imaging via superoscillation focusing of a radially polarized beam. *Optica* **5**, 86-92, doi:10.1364/OPTICA.5.000086 (2018).

16  Ren, H. *et al.* Complex-amplitude metasurface-based orbital angular momentum holography in momentum space. *Nature Nanotechnology* **15**, 948-955, doi:10.1038/s41565-020-0768-4 (2020).

17  Fang, X., Ren, H. & Gu, M. Orbital angular momentum holography for high-security encryption. *Nature Photonics* **14**, 102-108, doi:10.1038/s41566-019-0560-x (2020).

18  Ren, H. *et al.* Metasurface orbital angular momentum holography. *Nature Communications* **10**, 2986, doi:10.1038/s41467-019-11030-1 (2019).

19  Rego, L. *et al.* Generation of extreme-ultraviolet beams with time-varying orbital angular momentum. *Science* **364**, eaaw9486 (2019).

20  Ni, J. C. *et al.* Three-dimensional chiral microstructures fabricated by structured optical vortices in isotropic material. *Light Sci. Appl.* **6**, e17011 (2017).

21  Maguid, E. *et al.* Photonic spin-controlled multifunctional shared-aperture antenna array. *Science* **352**, 1202, doi:10.1126/science.aaf3417 (2016).

22  Maguid, E. *et al.* Multifunctional interleaved geometric-phase dielectric metasurfaces. *Light: Science & Applications* **6**, e17027-e17027, doi:10.1038/lsa.2017.27 (2017).

23  Wang, J. *et al.* All-dielectric metasurface grating for on-chip multi-channel orbital angular momentum generation and detection. *Opt. Express* **27**, 18794-18802, doi:10.1364/OE.27.018794 (2019).

24  Huang, C. *et al.* Ultrafast control of vortex microlasers. *Science* **367**, 1018-1021 (2020).

25  Sroor, H. *et al.* High-purity orbital angular momentum states from a visible metasurface laser. *Nature Photonics* **14**, 498-503, doi:10.1038/s41566-020-0623-z (2020).

26  Zhang, Z. *et al.* Tunable topological charge vortex microlaser. *Science* **368**, 760, doi:10.1126/science.aba8996 (2020).

27  Zhang, Z. *et al.* Ultrafast control of fractional orbital angular momentum of microlaser emissions. *Light: Science & Applications* **9**, 179, doi:10.1038/s41377-020-00415-3 (2020).

28  Fedyanin, D. Y., Krasavin, A. V., Arsenin, A. V. & Zayats, A. V. Lasing at the nanoscale: coherent emission of surface plasmons by an electrically driven nanolaser. *Nanophotonics* **9**, 3965-3975, doi:https://doi.org/10.1515/nanoph-2020-0157 (2020).

29  Mei, S. *et al.* Flat Helical Nanosieves. *Advanced Functional Materials* **26**, 5255-5262, doi:10.1002/adfm.201601345 (2016).

30  Veksler, D. *et al.* Multiple Wavefront Shaping by Metasurface Based on Mixed Random Antenna Groups. *ACS Photonics* **2**, 661-667, doi:10.1021/acsphotonics.5b00113 (2015).

31  Kerber, R. M. *et al.* Interaction of an Archimedean spiral structure with orbital angular momentum light. *New Journal of Physics* **20**, 095005, doi:10.1088/1367-2630/aae105 (2018).

32  Mehmood, M. Q. *et al.* Broadband spin-controlled focusing via logarithmic-spiral nanoslits of varying width. *Laser & Photonics Reviews* **9**, 674-681, doi:10.1002/lpor.201500116 (2015).

33  Li, Z. *et al.* Generation of high-order optical vortices with asymmetrical pinhole plates under plane wave illumination. *Opt. Express* **21**, 15755-15764, doi:10.1364/OE.21.015755 (2013).

34  Vogel, H. A better way to construct the sunflower head. *Mathematical Biosciences* **44**, 179-189, doi:https://doi.org/10.1016/0025-5564(79)90080-4 (1979).

35  Trevino, J., Cao, H. & Dal Negro, L. Circularly Symmetric Light Scattering from Nanoplasmonic Spirals. *Nano Letters* **11**, 2008-2016, doi:10.1021/nl2003736 (2011).

36  Niu, K. *et al.* Linear and nonlinear spin-orbital coupling in golden-angle spiral quasicrystals. *Opt. Express* **28**, 334-344, doi:10.1364/OE.373957 (2020).

37  Mihailescu, M. Natural quasy-periodic binary structure with focusing property in near field diffraction pattern. *Opt. Express* **18**, 12526-12536, doi:10.1364/OE.18.012526 (2010).

38  Huang, K. *et al.* Spiniform phase-encoded metagratings entangling arbitrary rational-order orbital angular momentum. *Light: Science & Applications* **7**, 17156-17156, doi:10.1038/lsa.2017.156 (2018).

39  Yang, Y., Thirunavukkarasu, G., Babiker, M. & Yuan, J. Orbital-Angular-Momentum Mode Selection by Rotationally Symmetric Superposition of Chiral States with Application to Electron Vortex Beams. *Physical Review Letters* **119**, 094802, doi:10.1103/PhysRevLett.119.094802 (2017).

40  Huang, K. *et al.* Ultrahigh-capacity non-periodic photon sieves operating in visible light. *Nature Communications* **6**, 7059, doi:10.1038/ncomms8059 (2015).

41  Mei, S. *et al.* On-chip discrimination of orbital angular momentum of light with plasmonic nanoslits. *Nanoscale* **8**, 2227-2233, doi:10.1039/C5NR07374J (2016).

42  Molina-Terriza, G., Torres, J. P. & Torner, L. Management of the Angular Momentum of Light: Preparation of Photons in Multidimensional Vector States of Angular Momentum. *Physical Review Letters* **88**, 013601, doi:10.1103/PhysRevLett.88.013601 (2001).

43  Koshy, T. in *Fibonacci and Lucas Numbers with Applications*    16-50 (2017).

44  Kim, H. *et al.* Synthesis and Dynamic Switching of Surface Plasmon Vortices with Plasmonic Vortex Lens. *Nano Letters* **10**, 529-536, doi:10.1021/nl903380j (2010).

45  Tsai, W.-Y. *et al.* Twisted Surface Plasmons with Spin-Controlled Gold Surfaces. *Advanced Optical Materials* **7**, 1801060, doi:10.1002/adom.201801060 (2019).

46  Yang, Y. *et al.* Deuterogenic Plasmonic Vortices. *Nano Letters* **20**, 6774-6779, doi:10.1021/acs.nanolett.0c02699 (2020).

47  Kahl, P. *et al.* Normal-Incidence Photoemission Electron Microscopy (NI-PEEM) for Imaging Surface Plasmon Polaritons. *Plasmonics* **9**, 1401-1407, doi:10.1007/s11468-014-9756-6 (2014).

48	Spektor, G. *et al.* Revealing the subfemtosecond dynamics of orbital angular momentum in nanoplasmonic vortices. *Science* **355**, 1187, doi:10.1126/science.aaj1699 (2017).
49	Davis, T. J. *et al.* Subfemtosecond and Nanometer Plasmon Dynamics with Photoelectron Microscopy: Theory and Efficient Simulations. *ACS Photonics* **4**, 2461-2469, doi:10.1021/acsphotonics.7b00676 (2017).
50	Davis, T. J. *et al.* Ultrafast vector imaging of plasmonic skyrmion dynamics with deep subwavelength resolution. *Science* **368**, eaba6415, doi:10.1126/science.aba6415 (2020).
51	Spektor, G. *et al.* Mixing the Light Spin with Plasmon Orbit by Nonlinear Light-Matter Interaction in Gold. *Physical Review X* **9**, 021031, doi:10.1103/PhysRevX.9.021031 (2019).
52	Shaohua, D. *et al.* On-chip trans-dimensional plasmonic router. *Nanophotonics*, 20200078, doi:https://doi.org/10.1515/nanoph-2020-0078 (2020).
53	Kahl, P. *et al.* Direct observation of surface plasmon polariton propagation and interference by time-resolved imaging in normal-incidence two photon photoemission microscopy. *Plasmonics* **13**, 239-246 (2018).

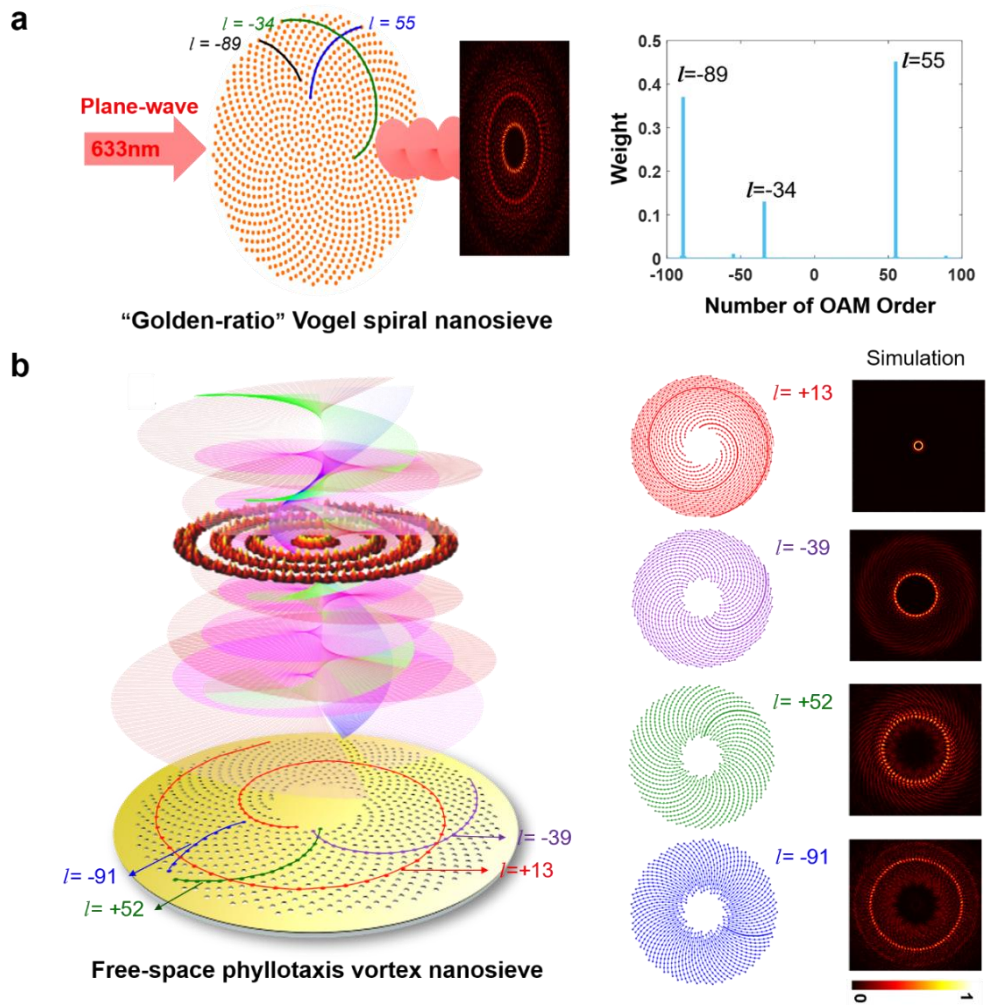

**Figure 1. Free-space phyllotaxis-inspired vortex nanosieve. a,** A "Golden-ratio" phyllotaxis pattern producing "Fibonacci" OAM series inspires us to design phyllotaxis-alike vortex nanosieves. In the Fresnel region, the diffractive pattern of a Vogel spiral nanosieve with 936 florets is simulated and mode coefficients are calculated. The corresponding results reveal the link between phyllotaxis patterns and optical vortices, thus inspiring us to design phyllotaxis-alike vortex nanosieves. **b,** Free-space phyllotaxis-inspired vortex nanosieve generating four different on-axis optical vortices simultaneously in free-space. Left panel: The red, purple, green, and blue solid circles denote the four different motifs concealed in the free-space phyllotaxis-inspired vortex nanosieve structure. Right panel: The four different sets of concealed spirals and their relevant simulated Fresnel diffraction intensity profiles.

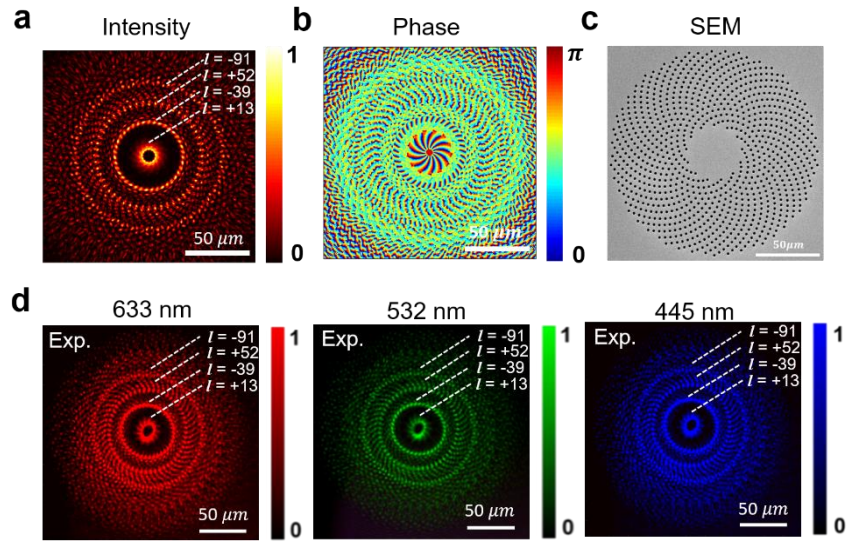

**Figure 2. Simulation and experimental results of the free-space phyllotaxis-inspired vortex nanosieve. a-b**, Simulated intensity (**a**) and phase (**b**) profiles of the free-space phyllotaxis-inspired vortex nanosieve. **c**, A top-view scanning electron microscope (SEM) image of the sample. **d**, The measured free-space Fresnel diffraction intensity profiles upon 663 nm, 532 nm, and 445 nm illumination, respectively.

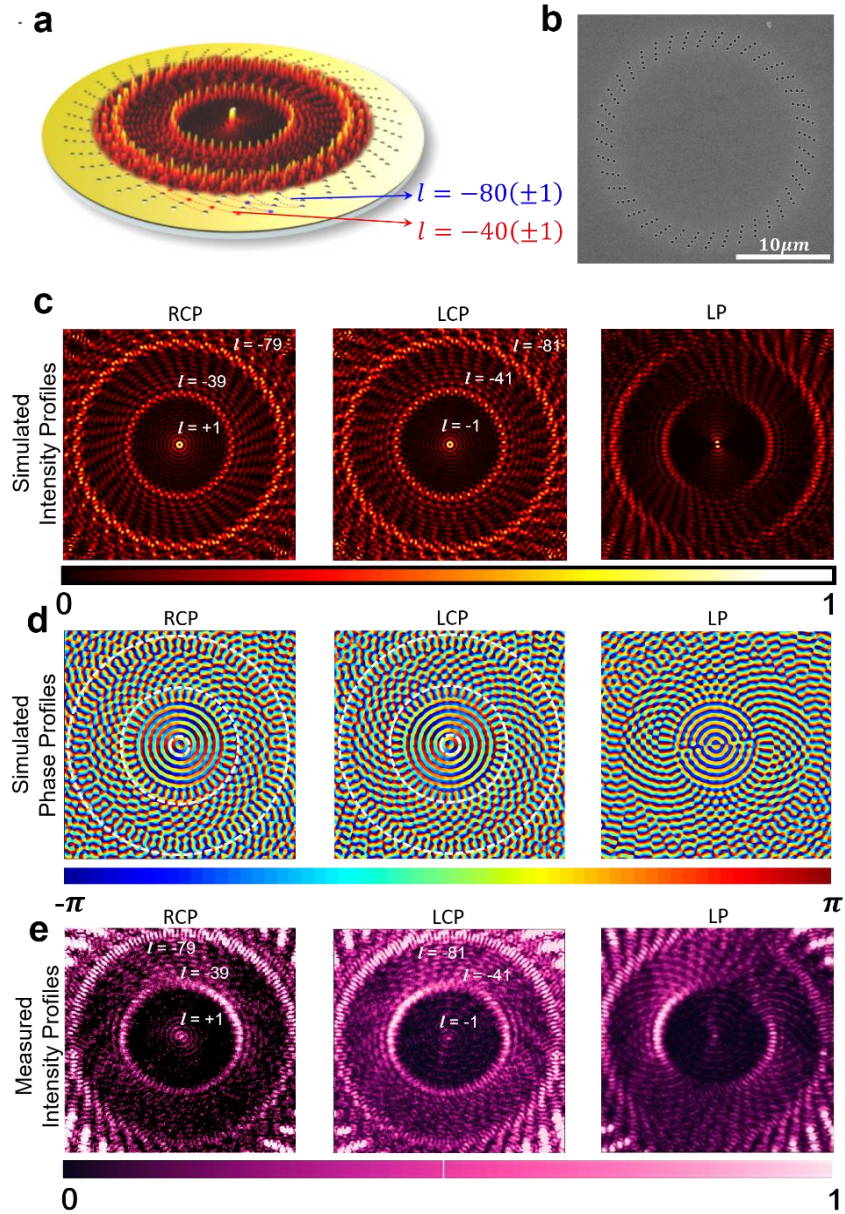

**Figure 3. Plasmonic phyllotaxis-inspired vortex nanosieve. a**, A plasmonic phyllotaxis-inspired vortex nanosieve which can yield three plasmonic vortex modes upon circular polarization states. The stereoscopic 'hot' map of the intensity distribution of the *z* component of the SPP (surface plasmon polariton) wave is depicted above the sample surface. The red and blue solid circles and dotted lines mark the trajectories of two hidden spiral patterns which can be encoded from the structure. The two sets of hidden spiral patterns can excite plasmonic vortices at l = 40($\pm$1), 80($\pm$1) upon circular polarizations. **b**, A top-view SEM image of the sample. The radius of each nanohole is 150 nm. **c-d**, The simulated intensity (**c**) and phase (**d**) profiles of generated plasmonic vortex modes excited by RCP, LCP, and LP illumination, respectively. **e**, Measured intensity profiles of the generated plasmonic vortex modes upon RCP, LCP and LP illumination.

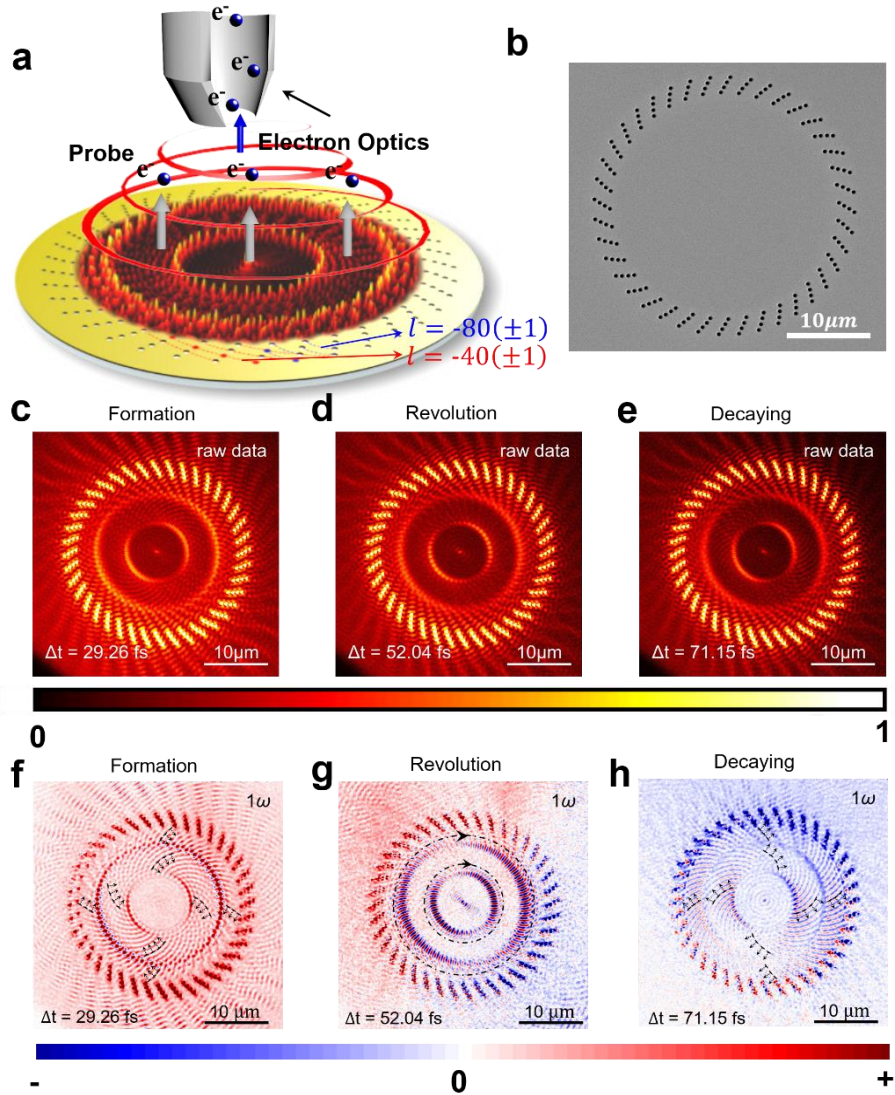

**Figure 4. Dynamic investigation of the plasmonic phyllotaxis-inspired vortex nanosieve.** **a**, During the dynamic investigation by TR-PEEM (time-resolved two-photon photoemission electron microscopy), the probe pulse interferes with the propagating SPPs and liberates photoelectrons in a two-photon photoemission process, which will then be imaged with the PEEM setup. **b**, A top-view SEM image of the sample. **c-e**, Raw experimental data of delay-time snapshots from measured TR-PEEM results. Three stages of vortex evolution upon RCP illumination can be identified: formation (**c**), revolution (**d**), and decay (**e**). All the data are shown on a logarithmic intensity scale. **f-g**, Processed images with only SPP dynamics that correspond to the delay times used in **c-e**. The small black dashed arrows indicate the propagation of vortex wavefronts in each stage. The black dashed arrows in panel **g** indicate the revolution direction of the vortex patterns.